# Voltage-Driven High-Speed Skyrmion Motion in a Skyrmion Shift Device

Yizheng Liu,[1] Na Lei,[1*] Chengxiang Wang,[1] Xichao Zhang,[2] Wang Kang,[1] Daoqian Zhu,[1] Yan Zhou,[2] Xiaoxi Liu,[3] Youguang Zhang,[1] Weisheng Zhao[1*]

[1] Fert Beijing Institute, BDBC, School of Electronic and Information Engineering, Beihang University, Beijing 100191, China
[2] School of Science and Engineering, The Chinese University of Hong Kong, Shenzhen 518172, China
[3] Department of Electrical and Computer Engineering, Shinshu University, 4-17-1 Wakasato, Nagano 380-8553, Japan

**ABSTRACT**

Magnetic skyrmions are promising information carriers for building future high-density and high-speed spintronic devices. However, to achieve a current-driven high-speed skyrmion motion, the required driving current density is usually very large, which could be energy inefficient and even destroy the device due to Joule heating. Here, we propose a voltage-driven skyrmion motion approach in a skyrmion shift device made of magnetic nanowires. The high-speed skyrmion motion is realized by utilizing the voltage shift, and the average skyrmion velocity reaches up to 259 m/s under 0.45 V applied voltage. In comparison with the widely studied vertical current-driven model, the energy dissipation is three orders of magnitude lower in our voltage-driven model, for the same speed motion of skyrmions. Our approach uncovers valuable opportunities for building skyrmion racetrack memories and logic devices with both ultra-low power consumption and ultra-high processing speed, which are appealing features for future spintronic applications.

Magnetic skyrmions show great potential as novel information carriers in spin memory and logic devices, because they have a number of merits including small size, low driving current density and topological stability.[1-7] Manipulations of magnetic skyrmions, including creation, motion and annihilation, have been intensively studied both theoretically[8-15] and experimentally.[16-25] Electric current is preferred to manipulate the skyrmion,[26-30] especially for skyrmion motion.[31-38] The threshold current density to drive skyrmions is around $10^6 \text{ A} \cdot \text{m}^{-2}$, which is over 5 orders of magnitude lower than that of conventional domain walls (DWs).[26] However, a large driving current density is required to achieve high-speed skyrmion motion for fast information processing, which would result in significant amount of Joule heating,[1,36] and induce instability of devices.[3,12] In addition, for current-driven skyrmion motion, geometric patterns[39,40] or additional electric gates[41] are usually attached to pin the skyrmion for realizing the addressable control, which increase the cost of operating energy and limit the maximum speed of skyrmion motion. For these reasons, more efficient and reliable means for controlling high-speed skyrmion motion are required.

Instead of current-driven approach, various approaches like spin wave,[9,19] magnetic field gradient,[14] and many more, have been proposed for driving skyrmion motion. Most of these methods lack practical convenience in integrated electrical circuit application and may not be energy efficient for commercial use.

The electric field or voltage has been proposed to be an energy-efficient method to manipulate magnetism[42,43] and has been progressively applied to the development of skyrmion-based applications.[25,44,45] Very recently, a voltage controlled magnetic anisotropy (VCMA) gradient model to drive skyrmions was numerically shown by Wang *et al.*.[46] Wherein, a wedged insulating layer is used to generate the magnetic anisotropy gradient, which can avoid the Joule heating effect. However, the wedged structure is not controllable after the layer is deposited. Thus, a simple structure with controllable and reliable method for high-speed skyrmion motion driven by voltage is highly demanded.

In this work, a series of equidistant identical electrodes on the nanowire with uniform thickness structure is proposed for high-speed skyrmion motion. A voltage-induced energy barrier and energy well on the sides of the skyrmion drive its directional motion to the addressable position along the nanowire. An analytical model is developed to describe the spin texture in skyrmion profile passing through an energy step induced by VCMA. To realize high-speed data delivery, parameters are optimized with and a solution is proposed for long-range skyrmion motion. With that optimization, the maximum velocity reaches 259 m/s when 0.45 V voltage is applied. Under 0.15 V voltage, only 8.4 aJ is consumed for one shift of a single skyrmion with an average velocity of 100 m/s. Our method shows great advantages in the potential application of skyrmion-based racetrack memory or spin logic devices.

## Model

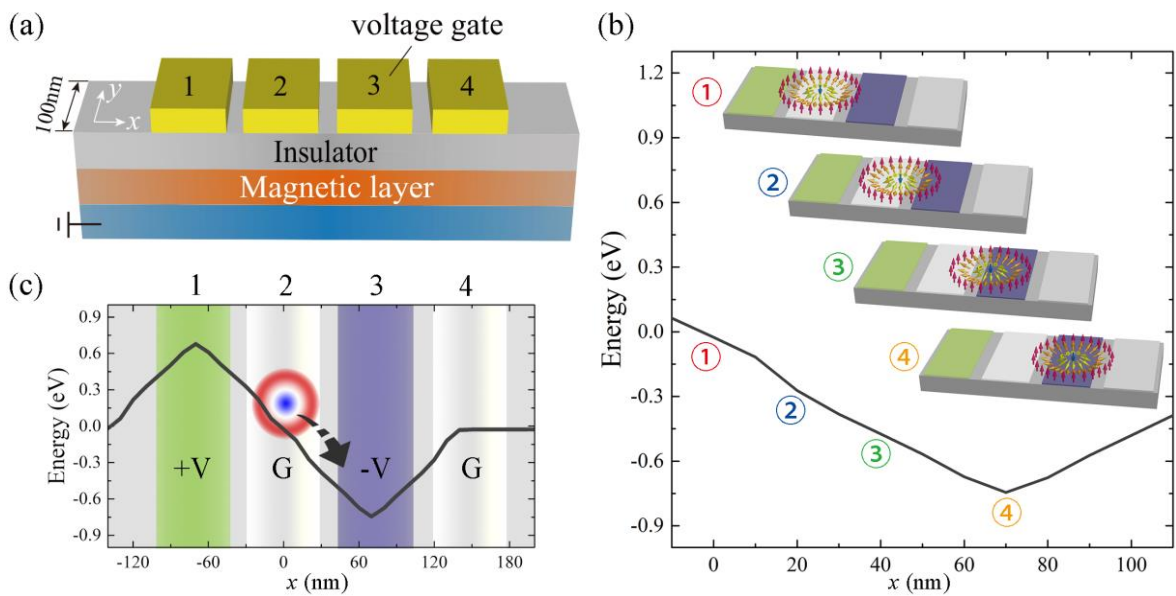

**Figure 1.** (a) Sketch of the proposed device structure. Uniform width contacts on a 100 nm wide nanowire serve as voltage gates, with equidistant spacing. The insulating layer acts as a dielectric layer for VCMA effect and the magnetic layer is for the generation of skyrmions. The nanowire is along the *x*-axis. (b) Different skyrmion locations and the corresponding total energy of the system. The energy profile evaluates continuously when skyrmions are located at different position along *x*-axis. (c) The variation of the total energy when a skyrmion is centered at different locations along a nanowire under voltages +V, ground (G), -V and ground (G) applied to contacts 1 to 4, successively. The potential barrier and well are formed at contacts 1 and 3, the skyrmion located at contact 2 falls into the energy well at contact 3. The potential well is of 0.98 eV when 0.3 V voltage is applied.

The device structure is shown in Fig. 1(a). Skyrmions move in a magnetic nanowire of 100 nm width, on which contacts with uniform width and equidistance are deposited as voltage gates. Numerical calculations are performed using the Object Oriented Micromagnetic Framework (OOMMF).

Due to the VCMA effect, voltages applied on contacts will induce local anisotropy change.[47-49] In our simulation, we set the VCMA coefficient ξ as 100 fJ/Vm, which has been experimentally demonstrated in Ir/CoFeB/MgO system[50] and Ta/CoFeB/MgO system.[51] The typical thickness of the MgO layer is 1 nm and applied voltage is assumed to be 0.3 V. Under these conditions, an electric field of 0.3 V/nm in the MgO layer induces a change of 30 kJ/m$^3$ in a magnetic anisotropy constant $K$, which is about 9.5% of an anisotropy change[52] (the typical dielectric breakdown field strength of MgO is about 2.4 V/nm[53,54]). In order to drive a skyrmion, a positive and a negative voltage are applied to contacts 1 and 3, respectively, while contacts 2 and 4 are connected to the ground. As shown in Fig. 1(b) and (c), the total energy evaluates continually when a skyrmion is located at different positions along the nanowire with voltage applied. An analytical model is developed to describe the energy change when different spin textures cross a step-like anisotropy profile, based on $E = -\boldsymbol{m} \cdot \boldsymbol{H}$ (See *Supplementary* I in *supporting information*, SI). The calculation results indicate that a spin texture with finite size will generate an energy gradient when it passes through a step-like anisotropy field induced by VCMA. Moreover, the energy gradient width is closely linked with the skyrmion size (See *Supplementary II*). In case the skyrmion radius $R_{sk}$ (radius of the $m_z = 0$ contour) is neglectable compared with the contact width, a step-like energy potential is expected.[46] In case of skyrmion radius being equivalent to the contact width, a continuous energy gradient is formed, shown as the black curve in Fig. 1 (c). The potential barrier and well of 0.98 eV on the left and right sides of a skyrmion centered at contact 2 are achieved. Then the skyrmion shifts to the contact 3, the potential well, realizing voltage-driven shift of magnetic skyrmion.

## Voltage-Driven Skyrmions Dynamics

For the analysis of skyrmion dynamics, Thiele equation is generally used with an assumption of rigid skyrmions. However, in our case, the skyrmion deformation is non-negligible, thus the Thiele equation can just serve for qualitative analysis. In consideration of the driving force from the energy gradient $F_g$ and repulsive force from the edge $F_e$, the equation is written as[28,55,56]

$$\boldsymbol{G} \times \boldsymbol{v} - \alpha \boldsymbol{D} \cdot \boldsymbol{v} + \boldsymbol{F}_g + \boldsymbol{F}_e = 0 \tag{2}$$

where $v$ is the instantaneous velocity of the skyrmion. $\boldsymbol{G}$ is the total gyromagnetic coupling vector, which only has a z-component $G_z = \int_{sk} dr^2 \boldsymbol{m} \cdot (\partial_x \boldsymbol{m} \times \partial_y \boldsymbol{m}) = 4\pi N_{sk}$ ($N_{sk}$ is the skyrmion number, and sk stands for the skyrmion region). $\boldsymbol{D}$ is the dissipative tensor, whose elements are given by $D_0 = D_{xx} = D_{yy} = \int_{sk} dr^2 \partial_x \boldsymbol{m} \cdot \partial_y \boldsymbol{m}$. $\boldsymbol{F}_g$ is only the function of $x$, which can be expressed as $F_g(x) = \partial_x E_{ani}(x)$ (See *Supplementary* I). $\boldsymbol{F}_e$ is perpendicular to the nanowire edge, thus it is along the y-axis. The analytical solution for the instantaneous velocity can be expressed as

$$\begin{pmatrix} v_x \\ v_y \end{pmatrix} = \frac{1}{G^2 + \alpha^2 D_0^2} \begin{pmatrix} F_g \alpha D_0 - F_e G \\ F_g G + F_e \alpha D_0 \end{pmatrix} \tag{3}$$

where $v_x$ and $v_y$ are the longitudinal and transverse components of instantaneous velocity, respectively.

As shown in Fig. 2(a), a skyrmion is driven towards the right contact (energy well) when the voltage pulses are applied. The skyrmion is not moving in a straight line along the nanowire as its trajectory (black dashed line) exhibits a noticeable transverse drift in voltage-driven motion. The corresponding longitudinal and transverse components $v_x$ and $v_y$ are plotted in Fig. 2(b) respectively, where $v_x$ is much larger than $v_y$. The voltage driving force and the edge repulsive force contribute to both $v_x$ and $v_y$, and both $v_x$ and $v_y$ are inversely proportional to the damping coefficient alpha. The transverse velocity $v_y$ causes the skyrmion Hall effect, resulting in a transverse drift in the voltage-driven motion of a skyrmion. To avoid the annihilation of skyrmions, the anisotropic constant $K$ is set higher at the edge.[57]

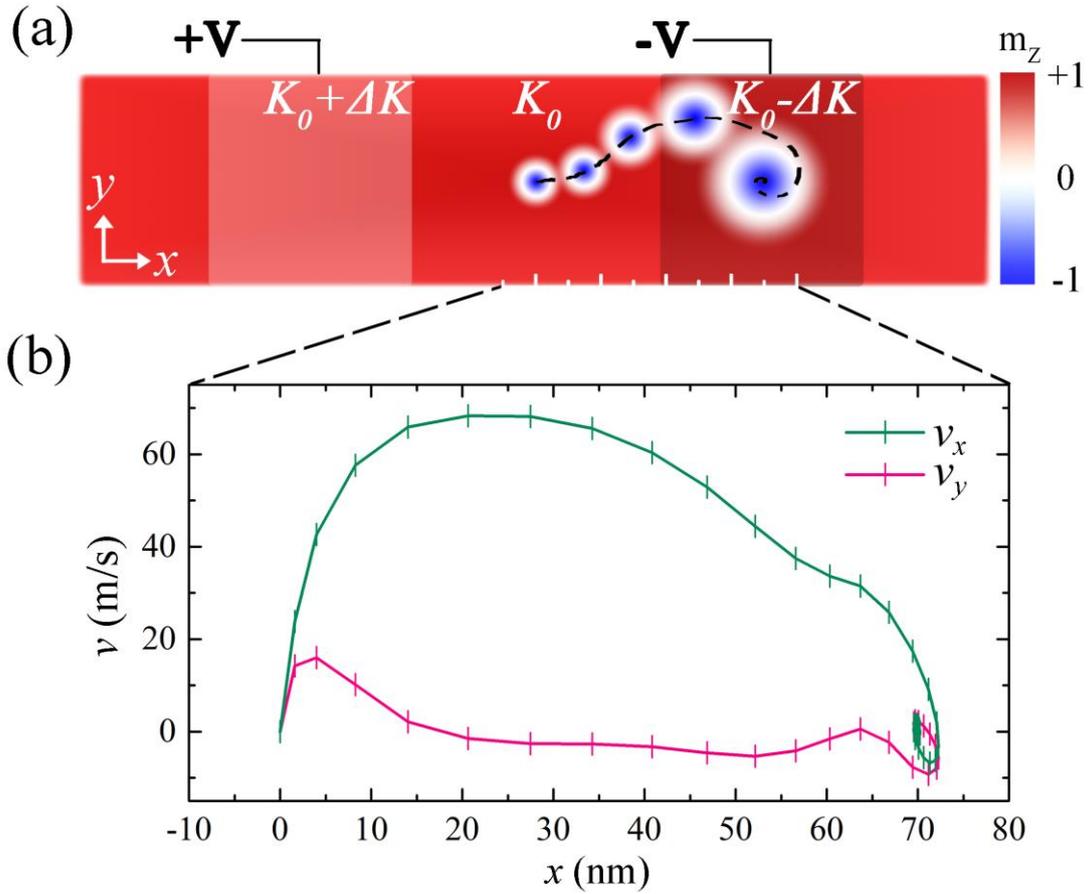

**Figure 2.** (a) Top-view of the nanowire and typical trajectory of the skyrmion shift. +V and -V are applied to the left and the right contacts to create a positive and a negative anisotropy constant change $\Delta K$. The regions with red color (blue color) represents $m_Z$ =+1 ($m_Z$ =-1) of the magnetic layer. The skyrmion size increases at different locations during one shift. (b) Evolution of instantaneous velocity during the shift. $v_x$ and $v_y$ are the longitudinal and transverse components of velocity obtained with anisotropic constant $K = 360 \text{ kJ/m}^3$ and Dzyaloshinskii-Moriya interaction constant $D = 1.2 \text{ mJ/m}^2$.

Furthermore, an obvious skyrmion size change can be observed during its shift, which is consistent with the theoretical expression of skyrmion radius[8]

$$R_{sk} \approx \frac{\Delta}{\sqrt{2(1-D/D_c)}} \qquad (4)$$

where $\Delta = \sqrt{A/K}$ and $D_c = 4\sqrt{AK}/\pi$. In our model, as the energy well is centered at the zone with decreased $K$, the skyrmion radius should increase when it moves towards the energy well.

One important feature of the voltage-driven motion is that, the skyrmion will always overshoot the center of an energy well and process for a while. The procession is very sensitive to the damping constant $\alpha$, and becomes extremely time-consuming at lower $\alpha$. It occupies only 0.05 ns in a 1.5 ns shift under $\alpha = 0.3$, but 14 ns in one 15 ns shift with $\alpha = 0.02$. This leads to a large reduction in the average velocity for a single shift.

## Long-Range Skyrmion Motion

In the skyrmion-based memory and logic devices, long-range skyrmion motion is demanded. To enhance the efficiency in long-range skyrmion motion, the relaxation process owing to the overshoot can be suppressed or avoided. One solution is to switch the voltages once skyrmions cross the center of the contacts, so that the procession is skipped and skyrmions continue to move to the designated contact directly (See Fig. 3).

As a demonstration, one motion containing 4 single shifts in Fig. 3 exhibits the behavior of long-range skyrmion motion driven by voltage. A long-range motion can be divided into 3 steps. That is, the first shift from $t_0$ to $t_1$, second from $t_1$ to $t_3$, and the last shift from $t_3$ to $t_4$. The average velocities of each shift follow the relationship $v_1 \approx v_2 > v_0 > v_3$ (shifts in the middle steps exhibit a similar velocity). For example, with parameters given in *method*, the simulation results are $v_0 = 46.9$ m/s, $v_1 = v_2 = 50.0$ m/s, $v_3 = 24.1$ m/s, in Fig. 3 and 20.0 m/s for one single shift in Fig. 2. Benefitting from this strategy, the average velocity in long-range motion is increased to 42.75 m/s, which is two times larger than that of a single shift. In a shift device, the average velocity of long-range skyrmion motion is improved due to this improvement. In the following text, $v_0$ stands for the average velocity of the first shift in a long-range motion, and the maximum limit of long-range motion average velocity is defined as $v_L$. $v_L$ equals to $v_1$ ($v_2$) when the number of shifts increases.

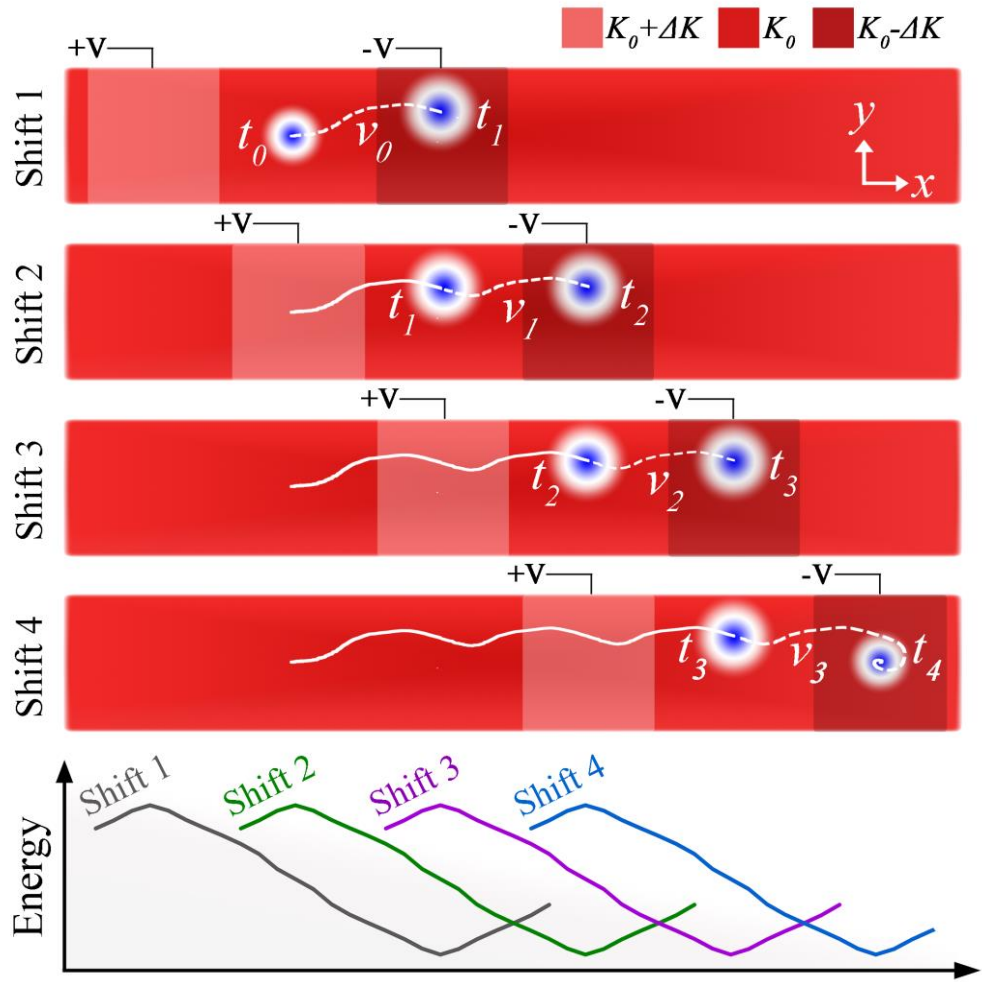

**Figure 3.** Long-range motion trajectory and corresponding energy profiles for each shift. Energy wells move from left to right as the applied voltages switch from shift 1 to shift 4. To avoid the time-consuming procession, voltages are switched at the moments once skyrmions pass the center of the contacts. These moments are marked as $t_1$ to $t_4$. $v_0$, $v_1$, $v_2$ and $v_3$ are the average velocity of each shift, separately. For the last shift, voltage is on until the skyrmion is stabilized.

## Velocity Optimization

Magnetic parameters of $K$ and $D$ are tuned to optimize the performance of the voltage-driven skyrmion motion. In our simulation, the transverse drifts are strongly suppressed with the increase in $D$ and a decrease in $K$. According to Equation (4), the radii of skyrmions increase with the increase in $D$ and decrease in $K$. Thus, the trajectory of the skyrmion motion strongly depends on its radius.

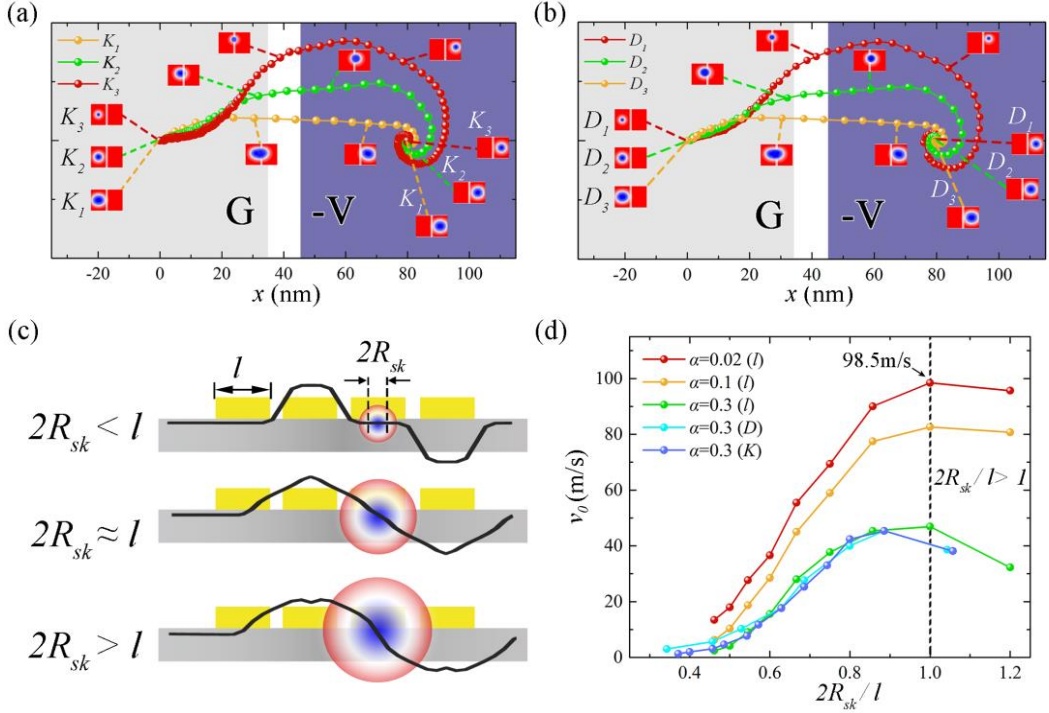

**Figure 4.** (a) Skyrmion trajectories with different $K$, and $D$. $K_1 = 360$ kJ/m$^3$, $K_2 = 335$ kJ/m$^3$ and $K_3 = 310$ kJ/m$^3$. $D = 1.2$ mJ/m$^2$. (b) Skyrmion shift trajectories with different $D$. $D_1 = 0.9$ mJ/m$^2$, $D_2 = 1.05$ mJ/m$^2$ and $D_3 = 1.2$ mJ/m$^2$. $K = 310$ kJ/m$^3$. The shape and size of skyrmions at different moments during the shifts are simulated and presented by the insets in panels (a) and (b). The applied voltage for the left and right parts are ground and -V, respectively. (c) Side-view and different energy profiles when skyrmion radius ($R_{sk}$) changes with a fixed contact width $l$. Energy platforms appear when $R_{sk}$ is much less than $l/2$, and it disappears when $R_{sk} \approx l/2$. (d) $v_0$ vs. $2R_{sk}/l$ with different $l$. To realize the $2R_{sk}/l$ modulation, we change $l$, $D$ or $K$ (See *Supplementary* IV).

The voltage-driven skyrmion motion is benefited from the formation of energy gradient when the non-uniform spin texture of a skyrmion crosses the boundaries of contacts. To realize the energy gradient, the ratio of skyrmion radius and the contact width should be sufficiently large, as shown in Fig. 4(c). For a skyrmion radius much smaller than the contact width, an energy platform appears at the initial skyrmion position and disables the skyrmion motion. As the skyrmion size expands, the energy platform shrinks until it finally disappears and the skyrmion starts to move when its spin texture is partially located on the energy gradient. With further expansion, the skyrmion size becomes larger than the contact width and hence the energy platform shows up again. For the purpose of velocity optimization, we vary the skyrmion radius till it is half of the contact width, and the ratio $2R_{sk}/l$ is induced. By manipulating the device parameters $D$, $K$ and $l$, the initial average velocity $v_0$ increases until $2R_{sk}/l$ reaches 1, as shown in Fig. 4(d). As long as the damping constant $\alpha$ and the ratio $2R_{sk}/l$ are fixed, $v_0$ is roughly constant regardless of the value of $D$, $K$ and $l$, and the maximum $v_0$ is obtained when $2R_{sk}/l = 1$. This conclusion is verified by a developed analytical model to find the optimal contact width, where the contribution from the edge is ignored (See *Supplementary III*). For skyrmions with 31 nm radius, the optimal contact width is found to be 58 nm, which is consistent with the OOMMF simulation results above.

By decreasing the damping constant, the velocity increases, which is consistent with equation (3). For 1 nm CoFeB layer, the typical damping constant varies in a wide range between 0.004~0.15 with different growth conditions and annealing treatments.[58,59] In our simulation, $v_0$ can reach 98.5 m/s under 0.3 V with $\alpha = 0.02$. Hence, higher velocity can be expected with further decrease in $\alpha$.

In our designed structure for the voltage-driven skyrmion motion, the optimal velocity $v_0$ can be achieved when the skyrmion size is equivalent to the contact width. Unlike $v_0$, the skyrmion shape in the initial stage is irregular for $v_L$. Thus, $v_L$ does not share the simple correlation with $2R_{sk}/l$ compared to that of $v_0$.

Under the optimization of $2R_{sk}/l = 1$, the average velocity in the initial stage and long range skyrmion motion are all simulated with different applied voltages and damping constants, as shown in Fig. 5(a) and (b).

The average velocity increases with the increase in applied voltages and the decrease in damping constants. Nevertheless, a large applied voltage induces the skyrmion deformation especially with low damping constant as indicated in the insets of Fig. 5(a) and (b). The simulation data are marked as open dots when the large skyrmion deformation exceeds the contact size, which is not in consideration for effective data delivery. Therefore, for $\alpha = 0.02$, the maximum $v_0$ is limited to 98.5 m/s with 0.3 V and the maximum $v_L$ is 259.3 m/s with 0.45 V. For the long-range motion under 0.45 V, one shift costs only 0.27 ns, leading to a switching frequency of 3.7 GHz.

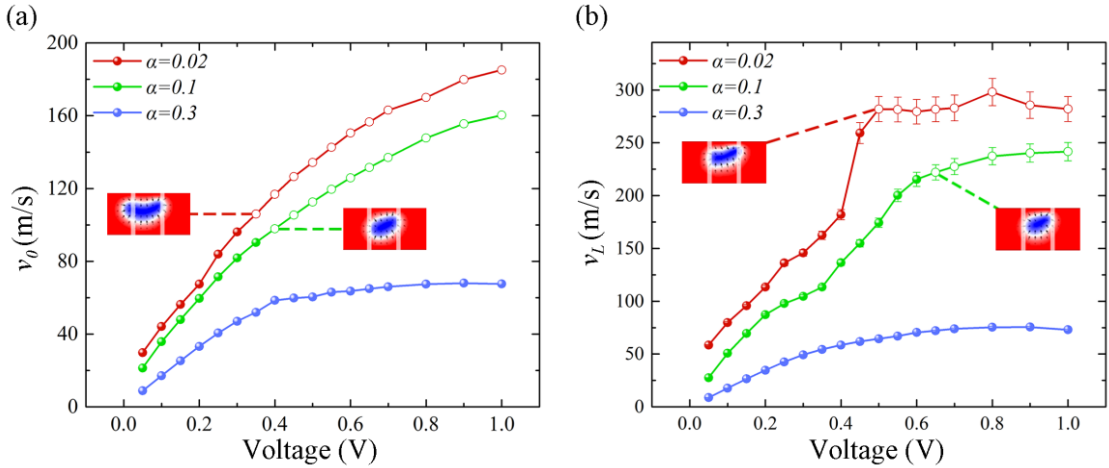

**Figure 5.** (a) $v_0$ and (b) $v_L$ vs. applied voltage V under different $\alpha$. Inserted images are simulated results of skyrmions indicating their size and shape at the moments when the center of skyrmions passes the center of the contacts. For open symbols, the skyrmion diameter is calculated to be larger than contact with, i.e., $2R_{sk}/l > 1$, which is inapplicable to our device.

The energy cost is estimated in our designed structure, which is simply expressed as $E = 0.5CV^2$ for a single contact in one operation, where $C$ is the device capacitance. When the contact width is fixed at 60 nm, $C$ is set to be 0.37 fF, and the energy consumption is 8.4 aJ per operation at 100 m/s velocity. To drive skyrmion motion under the same velocity, the voltage-driven method consumes three orders lower energy than that of the current-driven method (See *Supplementary V*).[37]

**Conclusion**

In this work we propose a voltage-driven skyrmion motion method and numerically demonstrate its feasibility. Parameters are optimized with a model describing the skyrmion profile. In our system, the ratio of skyrmion radius and voltage contact width is found to have a great impact on skyrmion velocity, which reaches its maximum value when the ratio equals to 1. To overcome the problem of overshooting and relaxation in long-range skyrmion motion, we come up with a solution, raising the long-range velocity to 259 m/s with 0.45 V. Our concept can be applied directly to the racetrack memory, with excellent characteristics like low-power consumption, high storage density and high switching frequency. In contrast to the current-driven method, the energy efficiency and addressable motion of voltage-driven method show great advantages in the application of skyrmion-based spintronic applications.

## METHODS

Numerical calculations were performed using the Object Oriented Micromagnetic Framework (OOMMF) which allows magnetization dynamics simulation by solving the Landau-Lifshitz-Gilbert equation, written as

$$\frac{\partial \boldsymbol{m}}{\partial t} = -\gamma_0 (\boldsymbol{m} \times h_{\text{eff}}) + \alpha \left( \boldsymbol{m} \times \frac{\partial \boldsymbol{m}}{\partial t} \right) \quad (1)$$

where $\boldsymbol{m}$ is the reduced magnetization, $\gamma_0$ is the gyromagnetic ratio and $h_{eff}$ is the effective field. Magnetic parameters based on typical CoFeB-MgO perpendicular magnetic anisotropy (PMA) systems are used to model the magnetic nanowire[60,61]: the PMA constant $K = 310 \sim 360 \text{ kJ/m}^3$, the Dzyaloshinskii-Moriya interaction (DMI) constant $D = 0.9 \sim 1.2 \text{ mJ/m}^2$, the saturation magnetization $M_s = 6 \times 10^5$ A/m, the exchange coefficient $A_{ex} = 1.12 \times 10^{-11}$ J/m, and the Gilbert damping constant $\alpha = 0.3, 0.1$ or $0.02$. To avoid the annihilation of skyrmions, anisotropic constant $K$ is set higher at the edge, namely $K_{edge} = 3 \text{ MJ/m}^3$.

## SUPPORTING INFORMATION

The supporting information contains: Energy evaluation with skyrmion motion, Energy profile change with different skyrmion radius, Optimization of average shift velocity, radius and velocity change with different magnetic parameters, Calculation of energy consumption,
and two videos for the voltage-driven motion under different velocity.

## Corresponding Author

Na Lei (na.lei@buaa.edu.cn)

Weisheng Zhao (weisheng.zhao@buaa.edu.cn)

## ACKNOWLEDGEMENTS

The authors gratefully acknowledge the National Natural Science Foundation of China (Grant No. 11574018, 61627813, 61571023, 11574137), the International Collaboration Project B16001, and the National Key Technology Program of China 2017ZX01032101 for their financial support of this work. X.Z. was supported by JSPS RONPAKU (Dissertation Ph.D.) Program. Y.Z. acknowledges the support by the President's Fund of CUHKSZ, and Shenzhen Fundamental Research Fund (Grant No. JCYJ20160331164412545 and JCYJ20170410171958839).

## REFERENCES

1. Fert, A.; Cros, V.; Sampaio, J. Skyrmions on the Track. *Nat. Nanotechnol*. **2013**, *8*, 152–156.
2. Tomasello, R.; Martinez, E.; Zivieri, R.; Torres, L.; Carpentieri, M.; Finocchio, G. A Strategy for the Design of Skyrmion Racetrack Memories. *Sci. Rep.* **2015**, *4*, 6784–6790.
3. Koshibae, W.; Kaneko, Y.; Iwasaki, J.; Kawasaki, M.; Tokura, Y.; Nagaosa, N. Memory Functions of Magnetic Skyrmions. *Jpn. J. Appl. Phys.* **2015**, *54*, 53001–53020.
4. Yu, G.; Upadhyaya, P.; Shao, Q.; Wu, H.; Yin, G.; Li, X.; He, C.; Jiang, W.; Han, X.; Amiri, P. K.; et al. Room-Temperature Skyrmion Shift Device for Memory Application. *Nano Lett.* **2017**, *17*, 261–268.

5. Zhou, Y.; Ezawa, M. A Reversible Conversion between a Skyrmion and a Domain-Wall Pair in a Junction Geometry. *Nat. Commun.* **2014**, *5*, 5652–5659.

6. Kang, W.; Huang, Y.; Zhang, X.; Zhou, Y.; Zhao, W. Skyrmion-Electronics: An Overview and Outlook. *Proc. IEEE* **2016**, *104*, 2040–2061.

7. Jiang, W.; Zhang, W.; Yu, G.; Jungfleisch, M. B.; Upadhyaya, P.; Somaily, H.; Pearson, J. E.; Tserkovnyak, Y.; Wang, K. L.; Heinonen, O.; *et al.* Mobile Néel Skyrmions at Room Temperature: Status and Future. *AIP Adv.* **2016**, *6*, 55602–55607.

8. Rohart, S.; Thiaville, A. Skyrmion Confinement in Ultrathin Film Nanostructures in the Presence of Dzyaloshinskii-Moriya Interaction. *Phys. Rev. B* **2013**, *88*, 184422–184429.

9. Mochizuki, M. Spin-Wave Modes and Their Intense Excitation Effects in Skyrmion Crystals. *Phys. Rev. Lett.* **2012**, *108*, 17601–17605.

10. Sun, L.; Cao, R. X.; Miao, B. F.; Feng, Z.; You, B.; Wu, D.; Zhang, W.; Hu, A.; Ding, H. F. Creating an Artificial Two-Dimensional Skyrmion Crystal by Nanopatterning. *Phys. Rev. Lett.* **2013**, *110*, 167201–167205.

11. Kong, L.; Zang, J. Dynamics of an Insulating Skyrmion under a Temperature Gradient. *Phys. Rev. Lett.* **2013**, *111*, 67203–67207.

12. Koshibae, W.; Nagaosa, N. Creation of Skyrmions and Antiskyrmions by Local Heating. *Nat. Commun.* **2014**, *5*, 5148–5158.

13. Li, Z.; Zhang, Y.; Huang, Y.; Wang, C.; Zhang, X.; Liu, Y.; Zhou, Y.; Kang, W.; Koli, S. C.; Lei, N. Strain-Controlled Skyrmion Creation and Propagation in Ferroelectric/ferromagnetic Hybrid Wires. *J. Magn. Magn. Mater.* **2018**, *455*, 19–24.

14. Wang, C.; Xiao, D.; Chen, X.; Zhou, Y.; Liu, Y. Manipulating and Trapping Skyrmions by Magnetic Field Gradients. *New J. Phys.* **2017**, *19*, 083008–082015.

15. Huang, Y.; Kang, W.; Zhang, X.; Zhou, Y.; Zhao, W. Magnetic Skyrmion-Based Synaptic Devices. *Nanotechnology* **2017**, *28*, 08LT02.

16. Seki, S.; Yu, X. Z.; Ishiwata, S.; Tokura, Y. Observation of Skyrmions in a Multiferroic Material. *Science* **2012**, *336*, 198–201.

17. Boulle, O.; Vogel, J.; Yang, H.; Pizzini, S.; de Souza Chaves, D.; Locatelli, A.; Menteş, T. O.; Sala, A.; Buda-Prejbeanu, L. D.; Klein, O.; et al. Room-Temperature Chiral Magnetic Skyrmions in Ultrathin Magnetic Nanostructures. *Nat. Nanotechnol.* **2016**, *11*, 449–454.

18. Onose, Y.; Okamura, Y.; Seki, S.; Ishiwata, S.; Tokura, Y. Observation of Magnetic Excitations of Skyrmion Crystal in a Helimagnetic Insulator $Cu_2OSeO_3$. *Phys. Rev. Lett.* **2012**, *109*, 37603–37607.

19. Mochizuki, M.; Yu, X. Z.; Seki, S.; Kanazawa, N.; Koshibae, W.; Zang, J.; Mostovoy, M.; Tokura, Y.; Nagaosa, N. Thermally Driven Ratchet Motion of a Skyrmion Microcrystal and Topological Magnon Hall Effect. *Nat. Mater.* **2014**, *13*, 241–246.

20. Shibata, K.; Yu, X. Z.; Hara, T.; Morikawa, D.; Kanazawa, N.; Kimoto, K.; Ishiwata, S.; Matsui, Y.; Tokura, Y. Towards Control of the Size and Helicity of Skyrmions in Helimagnetic Alloys by Spin–orbit Coupling. *Nat. Nanotechnol.* **2013**, *8*, 723–728.

21. Romming, N.; Kubetzka, A.; Hanneken, C.; von Bergmann, K.; Wiesendanger, R. Field-Dependent Size and Shape of Single Magnetic Skyrmions. *Phys. Rev. Lett.* **2015**, *114*, 177203–177207.

22. White, J. S.; Levatić, I.; Omrani, A. A.; Egetenmeyer, N.; Prša, K.; Živković, I.; Gavilano, J. L.; Kohlbrecher, J.; Bartkowiak, M.; Berger, H.; et al. Electric Field Control of the Skyrmion Lattice in $Cu_2OSeO_3$. *J. Phys. Condens. Matter* **2012**, *24*, 432201–432207.

23. Finazzi, M.; Savoini, M.; Khorsand, A. R.; Tsukamoto, A.; Itoh, A.; Duò, L.; Kirilyuk, A.; Rasing, T.; Ezawa, M. Laser-Induced Magnetic Nanostructures with Tunable Topological Properties. *Phys. Rev. Lett.* **2013**, *110*, 177205–177209.

24. Lin, S.-Z.; Batista, C. D.; Reichhardt, C.; Saxena, A. Ac Current Generation in Chiral Magnetic Insulators and Skyrmion Motion Induced by the Spin Seebeck Effect. *Phys. Rev. Lett.* **2014**, *112*, 187203–187207.

25. Schott, M.; Bernand-Mantel, A.; Ranno, L.; Pizzini, S.; Vogel, J.; Béa, H.; Baraduc, C.; Auffret, S.; Gaudin, G.; Givord, D. The Skyrmion Switch: Turning Magnetic Skyrmion Bubbles on and off with an Electric Field. *Nano Lett.* **2017**, *17*, 3006–3012.

26. Jonietz, F.; Muhlbauer, S.; Pfleiderer, C.; Neubauer, A.; Munzer, W.; Bauer, A.; Adams, T.; Georgii, R.; Boni, P.; Duine, R. A.; et al. Spin Transfer Torques in MnSi at Ultralow Current Densities. *Science* **2010**, *330*, 1648–1651.

27. Romming, N.; Hanneken, C.; Menzel, M.; Bickel, J. E.; Wolter, B.; Von Bergmann, K.; Kubetzka, A.; Wiesendanger, R. Writing and Deleting Single Magnetic Skyrmions. *Science* **2013**, *341*, 636–639.

28. Everschor, K.; Garst, M.; Binz, B.; Jonietz, F.; Mühlbauer, S.; Pfleiderer, C.; Rosch, A. Rotating Skyrmion Lattices by Spin Torques and Field or Temperature Gradients. *Phys. Rev. B* **2012**, *86*, 54432–54442.

29. Büttner, F.; Lemesh, I.; Schneider, M.; Pfau, B.; Günther, C. M.; Hessing, P.; Geilhufe, J.; Caretta, L.; Engel, D.; Krüger, B.; et al. Field-Free Deterministic Ultrafast Creation of Magnetic Skyrmions by Spin–orbit Torques. *Nat. Nanotechnol.* **2017**, *12*, 1040–1044.

30. Legrand, W.; Maccariello, D.; Reyren, N.; Garcia, K.; Moutafis, C.; Moreau-Luchaire, C.; Collin, S.; Bouzehouane, K.; Cros, V.; Fert, A. Room-Temperature Current-Induced Generation and Motion of Sub-100 nm Skyrmions. *Nano Lett*. **2017**, *17*, 2703–2712.

31. Iwasaki, J.; Mochizuki, M.; Nagaosa, N. Current-Induced Skyrmion Dynamics in Constricted Geometries. *Nat. Nanotechnol.* **2013**, *8*, 742–747.

32. Zang, J.; Mostovoy, M.; Han, J. H.; Nagaosa, N. Dynamics of Skyrmion Crystals in Metallic Thin Films. *Phys. Rev. Lett.* **2011**, *107*, 136804–136808.

33. Iwasaki, J.; Koshibae, W.; Nagaosa, N. Colossal Spin Transfer Torque Effect on Skyrmion along the Edge. *Nano Lett*. **2014**, *14*, 4432–4437.

34. Schulz, T.; Ritz, R.; Bauer, A.; Halder, M.; Wagner, M.; Franz, C.; Pfleiderer, C.; Everschor, K.; Garst, M.; Rosch, A. Emergent Electrodynamics of Skyrmions in a Chiral Magnet. *Nat. Phys*. **2012**, *8*, 301–304.

35. Yu, X. Z.; Kanazawa, N.; Zhang, W. Z.; Nagai, T.; Hara, T.; Kimoto, K.; Matsui, Y.; Onose, Y.; Tokura, Y. Skyrmion Flow near Room Temperature in an Ultralow Current Density. *Nat. Commun*. **2012**, *3*, 988–993.

36. Jiang, W.; Upadhyaya, P.; Zhang, W.; Yu, G.; Jungfleisch, M. B.; Fradin, F. Y.; Pearson, J. E.; Tserkovnyak, Y.; Wang, K. L.; Heinonen, O.; et al. Blowing Magnetic Skyrmion Bubbles. *Science* **2015**, *349*, 283–286.

37. Woo, S.; Litzius, K.; Krüger, B.; Im, M. Y.; Caretta, L.; Richter, K.; Mann, M.; Krone, A.; Reeve, R. M.; Weigand, M.; et al. Observation of Room-Temperature Magnetic Skyrmions and Their Current-Driven Dynamics in Ultrathin Metallic Ferromagnets. *Nat. Mater*. **2016**, *15*, 501–506.

38. Yu, G.; Upadhyaya, P.; Li, X.; Li, W.; Kim, S. K.; Fan, Y.; Wong, K. L.; Tserkovnyak, Y.; Amiri, P. K.; Wang, K. L. Room-Temperature Creation and Spin-Orbit Torque Manipulation of Skyrmions in Thin Films with Engineered Asymmetry. *Nano Lett.* **2016**, *16*, 1981–1988.

39. Fook, H. T.; Gan, W. L.; Lew, W. S. Gateable Skyrmion Transport via Field-Induced Potential Barrier Modulation. *Sci. Rep.* **2016**, *6*, 21099–21106.

40. Müller, J.; Rosch, A. Capturing of a Magnetic Skyrmion with a Hole. *Phys. Rev. B* **2015**, *91*, 54410–54419.

41. Kang, W.; Huang, Y.; Zheng, C.; Lv, W.; Lei, N.; Zhang, Y.; Zhang, X.; Zhou, Y.; Zhao, W. Voltage Controlled Magnetic Skyrmion Motion for Racetrack Memory. *Sci. Rep.* **2016**, *6*, 23164–23174.

42. Ohno, H.; Chiba, D.; Matsukura, F.; Omiya, T.; Abe, E.; Dietl, T.; Ohno, Y.; Ohtani, K. Electric-Field Control of Ferromagnetism. *Nature* **2000**, *408*, 944–946.

43. Ohno, H. A Window on the Future of Spintronics. *Nat. Mater.* **2010**, *9*, 952–954.

44. Hsu, P.-J.; Kubetzka, A.; Finco, A.; Romming, N.; von Bergmann, K.; Wiesendanger, R. Electric-Field-Driven Switching of Individual Magnetic Skyrmions. *Nat. Nanotechnol.* **2016**, *12*, 123–126.

45. Ma, C.; Zhang, X.; Yamada, Y.; Xia, J.; Ezawa, M.; Jiang, W.; Zhou, Y.; Morisako, A.; Liu, X. Creation and Directional Motion of Chiral Spin Textures Induced by Electric Fields. *ArXiv Preprint* **2017**, arXiv:1708.02023.

46. Wang, X.; Gan, W. L.; Martinez, J. C.; Tan, F. N.; Jalil, M. B. A.; Lew, W. S. Efficient Skyrmion Transport Mediated by a Voltage Controlled Magnetic Anisotropy Gradient. *Nanoscale* **2018**, *10*, 733–740.

47. Wang, W. G.; Li, M.; Hageman, S.; Chien, C. L. Electric-Field-Assisted Switching in Magnetic Tunnel junctions. *Nat. Mater.* **2012**, *11*, 64–68.

48. Shiota, Y.; Nozaki, T.; Bonell, F.; Murakami, S.; Shinjo, T.; Suzuki, Y. Induction of Coherent Magnetization Switching in a Few Atomic Layers of FeCo Using Voltage Pulses. *Nat. Mater*. **2012**, *11*, 39–43.

49. Kanai, S.; Yamanouchi, M.; Ikeda, S.; Nakatani, Y.; Matsukura, F.; Ohno, H. Electric Field-Induced Magnetization Reversal in a Perpendicular-Anisotropy CoFeB-MgO Magnetic Tunnel Junction. *Appl. Phys. Lett.* **2012**, *101*, 122403–122405.

50. Skowroński, W.; Nozaki, T.; Shiota, Y.; Tamaru, S.; Yakushiji, K.; Kubota, H.; Fukushima, A.; Yuasa, S.; Suzuki, Y. Perpendicular Magnetic Anisotropy of Ir/CoFeB/MgO Trilayer System Tuned by Electric Fields. *Appl. Phys. Express* **2015**, *8*, 53003–53006.

51. Li, X.; Fitzell, K.; Wu, D.; Karaba, C. T.; Buditama, A.; Yu, G.; Wong, K. L.; Altieri, N.; Grezes, C.; Kioussis, N.; et al. Enhancement of Voltage-Controlled Magnetic Anisotropy through Precise Control of Mg Insertion Thickness at CoFeB|MgO Interface. *Appl. Phys. Lett.* **2017**, *110*, 52401–52405.

52. Peng, S.; Wang, M.; Yang, H.; Zeng, L.; Nan, J.; Zhou, J.; Zhang, Y.; Hallal, A.; Chshiev, M.; Wang, K. L.; *et al.* Origin of Interfacial Perpendicular Magnetic Anisotropy in MgO/CoFe/metallic Capping Layer Structures. *Sci. Rep.* **2016**, *5*, 18173–18177.

53. Dimitrov, D. V.; Gao, Z.; Wang, X.; Jung, W.; Lou, X.; Heinonen, O. G. Dielectric Breakdown of MgO Magnetic Tunnel Junctions. *Appl. Phys. Lett.* **2009**, *94*, 123110–123112.

54. Zhang, Y.; Lin, X.; Adam, J.-P.; Agnus, G.; Kang, W.; Cai, W.; Coudevylle, J.-R.; Isac, N.; Yang, J.; Yang, H.; *et al.* Heterogeneous Memristive Devices Enabled by Magnetic Tunnel Junction Nanopillars Surrounded by Resistive Silicon Switches. *Adv. Electron. Mater.* **2018**, 1700461–1700468.

55. Thiele, A. A. Steady-State Motion of Magnetic Domains. *Phys. Rev. Lett.* **1973**, *30*, 230–233.

56. Thiaville, A.; Nakatani, Y.; Miltat, J.; Suzuki, Y. Micromagnetic Understanding of Current-Driven Domain Wall Motion in Patterned Nanowires. *Europhys. Lett.* **2005**, *69*, 990–996.

57. Lai, P.; Zhao, G. P.; Tang, H.; Ran, N.; Wu, S. Q.; Xia, J.; Zhang, X.; Zhou, Y. An Improved Racetrack Structure for Transporting a Skyrmion. *Sci. Rep.* **2017**, *7*, 45330–45337.

58. Liu, X.; Zhang, W.; Carter, M. J.; Xiao, G. Ferromagnetic Resonance and Damping Properties of CoFeB Thin Films as Free Layers in MgO-Based Magnetic Tunnel Junctions. *J. Appl. Phys*. **2011**, *110*, 33910–33914.

59. Iihama, S.; Ma, Q.; Kubota, T.; Mizukami, S.; Ando, Y.; Miyazaki, T. Damping of Magnetization Precession in Perpendicularly Magnetized CoFeB Alloy Thin Films. *Appl. Phys. Express* **2012**, *5*, 83001–83003.

60. Ikeda, S.; Miura, K.; Yamamoto, H.; Mizunuma, K.; Gan, H. D.; Endo, M.; Kanai, S.; Hayakawa, J.; Matsukura, F.; Ohno, H. A Perpendicular-Anisotropy CoFeB-MgO Magnetic Tunnel Junction. *Nat. Mater.* **2010**, *9*, 721–724.

61. Torrejon, J.; Kim, J.; Sinha, J.; Mitani, S.; Hayashi, M.; Yamanouchi, M.; Ohno, H. Interface Control of the Magnetic Chirality in CoFeB/MgO Heterostructures with Heavy-Metal Underlayers. *Nat. Commun.* **2014**, *5*, 1–8.

## *Supporting Information for*
## Voltage-Driven High-Speed Skyrmion Motion in a Skyrmion Shift Device

**Supplementary I: Energy evaluation with skyrmion motion**

To calculate the evaluation of the total energy during skyrmion motion, we develop an analytical model to describe the energy change when different spin textures cross a step-like anisotropy profile, based on $E = -\boldsymbol{m} \cdot \boldsymbol{H}$. The anisotropic energy when the center of a domain structure arrives at $x$ can be expressed as

$$E_{ani}(x) = \int_\Omega -\boldsymbol{m}(x) \cdot \boldsymbol{H} d\Omega \qquad (S1)$$

where $\Omega$ stands for the integration interval (the whole nanowire) and $\boldsymbol{H}$ is the effective field considering the VCMA effect, who only has a *z*-component. The magnetic moment distribution in the nanowire is described by $\boldsymbol{m}(x)$, which changes with the spin texture motion. Once we know the analytical expression of the spin texture $\boldsymbol{m}(x)$, $E_{ani}(x)$ can be calculated numerically.

The energy evaluation during the motion for three different spin textures (uniform square domain, uniform circular domain, and skyrmion) is calculated respectively. The domain structure moves along the *y*-axis of a nanowire with finite width $l$ and infinite length. The boundary of the low-*K* region is set as $x = 0$, and the normalized analytical expression of $\boldsymbol{H}$ is

$$H_z(x) = \begin{cases} -1, & if\ x \geq 0 \\ 0, & if\ x < 0 \end{cases} \qquad (S2)$$

where $-1$ stands for the normalized effective field induced by VCMA effect.

Firstly, we start the calculation from a uniform square domain with side lengths of *a* (blue and red arrows indicate spin-down and spin-up respectively) is shown is Fig. S1. We let this domain structure move into the low-*K* region (dark gray region, stands for the VCMA effect). The position of this square domain is described by its center $(x_{rec}, y_{rec})$. The analytical expression of $\boldsymbol{m}(x_{rec})$ just has a $m_z$ component, written as

$$m_z(x_{rec}) = \begin{cases} -1, & if \ -\frac{a}{2} < x - x_{rec} < \frac{a}{2} \\ 1, & if \ else \end{cases} \quad (S3)$$

In this case, the energy evaluates linearly when the square spin-down structure crosses the boundary of the low-$K$ region. The width of linear decrease of the energy is equal to $a$, as shown in Fig. S1.

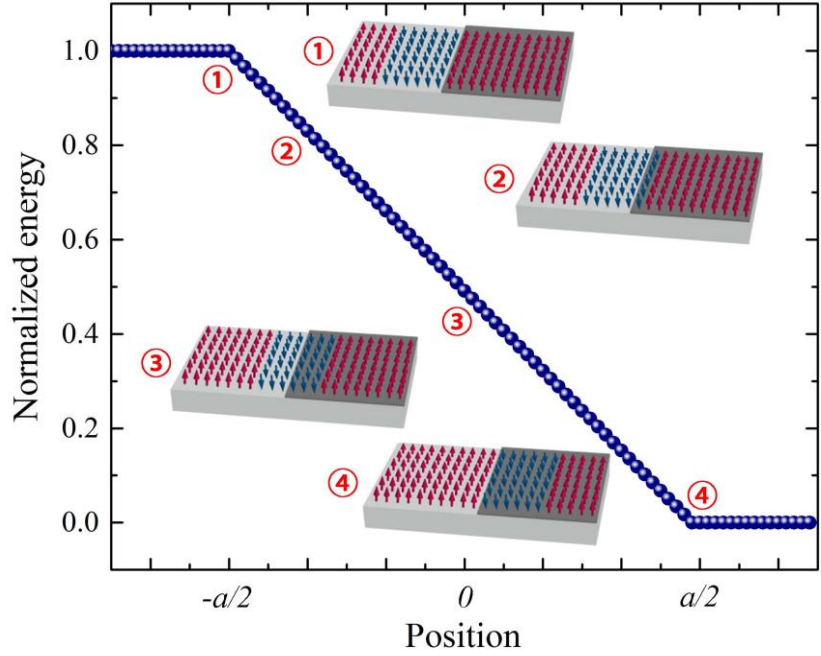

**Figure S1.** Normalized energy profile when a uniform square spin texture moves into a low-$K$ region (dark grey). Red and blue arrows stand for spin-up and spin-down moments, respectively.

Secondly, we simulate the energy evaluation when a uniform circular shape spin-down structure moves into the low-$K$ region. The position of this circular shape domain is described by its center $(x_r, y_r)$ and its radius is set as $R$. Similarly, the analytical expression of $\boldsymbol{m}(x_r)$ only has the $m_z$ component, written as

$$m_z(x_r) = \begin{cases} -1, & if \ (x - x_r)^2 + (y - y_r)^2 < R^2 \\ 1, & if \ else \end{cases} \quad (S4)$$

In this case, the energy evaluates continually, but the energy gradient behavior is different from that of the square spin-down structure, as shown in Fig. S2.

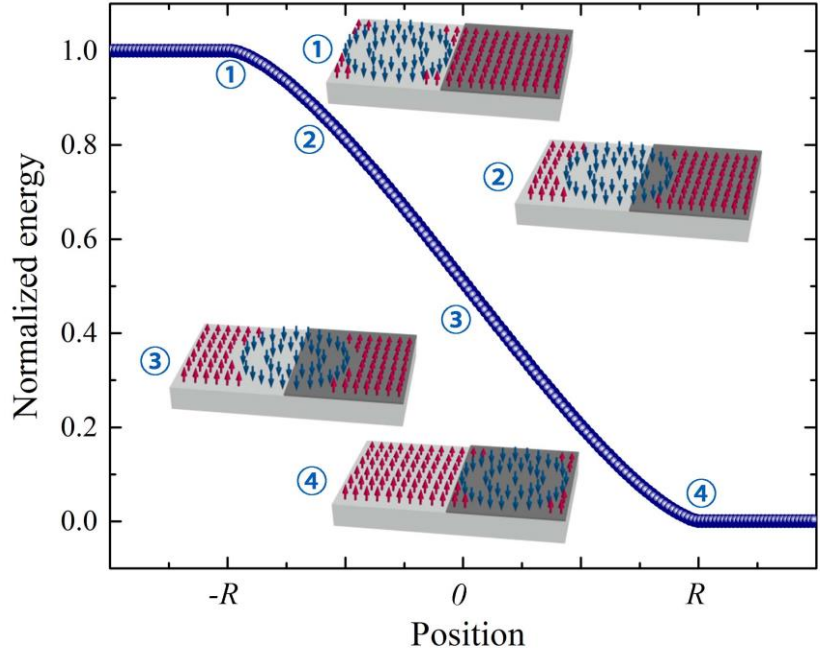

**Figure S2.** Normalized energy profile when a uniform circular shape spin-down moves into a low-*K* region (dark grey).

Finally, we simulate the energy evaluation when a skyrmion moves into the low-*K* region. The position of the skyrmion is described by its center $(x_{sk}, y_{sk})$ and its radius is set as $R_{sk}$ (radius of the $m_z = 0$ contour). An analytical model is induced to describe the skyrmion texture. In a spherical coordinate system, magnetization at $r$ is described by polar and azimuthal angles $\Theta(r, \phi)$ and $\Phi(r, \phi)$. For a skyrmion centered at $r = 0$, the polar angle can be expressed as[1,2]

$$\Theta(r) = 2 \arctan \left[ \frac{\sinh\left(\frac{R_{sk}}{w}\right)}{\sinh\left(\frac{r}{w}\right)} \right] \quad (S5)$$

where $w$ is the domain wall width. To apply this model in the rectangular coordinate system, the polar angle in the spherical coordinate $\Theta(r)$ is transformed to $\Theta(x, y)$ with the skyrmion centered at $(0, 0)$. The magnetic anisotropy energy when the skyrmion arrives at $(x_{sk}, y_{sk})$ can be expressed as

$$E_{ani}(x_{sk}) = \int -\boldsymbol{m}(x_{sk}) \cdot \boldsymbol{H} d\Omega$$

$$= \int -\cos(\Theta(x - x_{sk}, y - y_{sk})) \cdot |\boldsymbol{H}(x,y)| dxdy \qquad (S6)$$

In this case, the energy evaluates continually and the energy gradient is different with that of the circular shape spin-down domain, as shown in Fig. S3. The energy almost doesn't change at the beginning and the energy gradient appears when the $m_z = 0$ contour crosses the low-$K$ boundary.

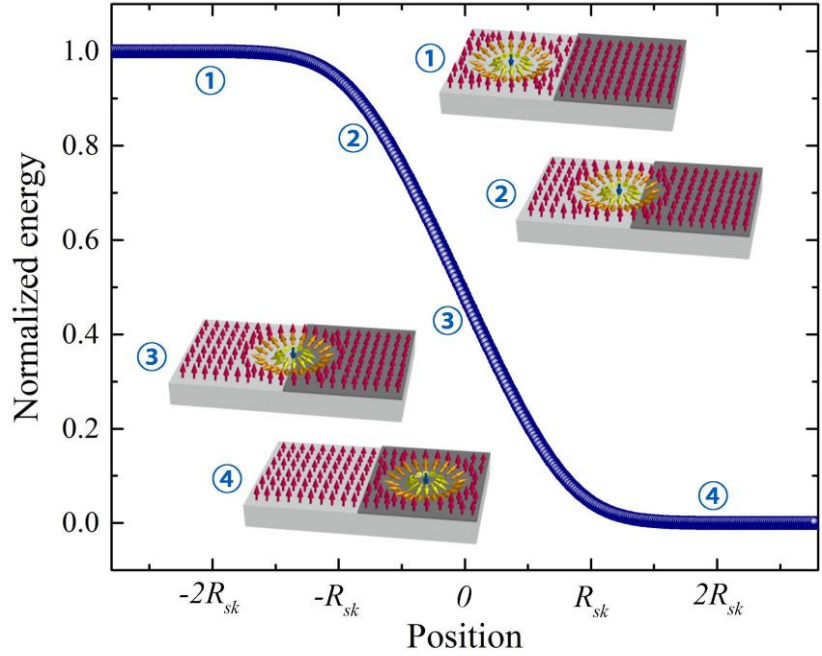

**Figure S3.** Normalized energy profile when a skyrmion moves into a low-$K$ region (Dark grey).

**Supplementary II: Energy profile change with different skyrmion radius**

In this part, we study the energy profile with varying the skyrmion radius $R_{sk}$. The voltage is applied as shown in Fig. 1 in the main text and the contact width $l$ is fixed at 60 nm. An energy platform appears in the energy profile when $2R_{sk} < l$, and it disappears when $2R_{sk} \approx l$ and reappears when $2R_{sk} > l$, as shown in Fig. S4. This change is consistent with the result in *Supplementary I* that the energy gradient width depends on the skyrmion radius.

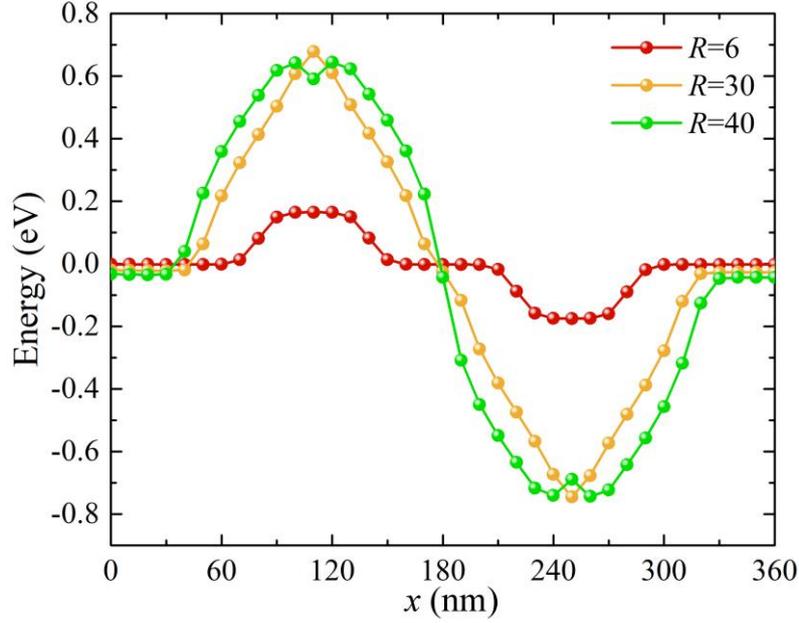

**Figure S5.** Energy profile change with different skyrmion radius. The contact width is fixed at 60 nm.

**Supplementary III: Optimization of average shift velocity**

According to equation (2) and (3) in the main text, the continuous change of $E_{ani}$ generates the driving force $F_g$ for the skyrmion motion, $F_g(x) = \partial_x E_{ani}(x)$. Assuming that the nanowire width is much larger than the skyrmion width, we ignore the force from the edge and the expression (3) for *x*-component can be simplified as

$$v_x(x) = \frac{F_{grad}(x)\alpha D_0}{G^2 + \alpha^2 D_0^{\,2}} \tag{S8}$$

We name the beginning and the destination of one skyrmion shift as $x_0$ and $x_1$. From $x$ to $x + dx$ in $[x_0, x_1]$, the time consumed can be estimated as $dt(x) = \frac{dx}{(v_x(x)+v_x(x+dx))/2}$. Therefore, the total time consumed in this shift is the integration of $dt(x)$ for $x$ in $[x_0, x_1]$, and the average velocity of this shift can be expressed as $v = \frac{x_1-x_0}{\int_{x_0}^{x_1} dt(x)}$. Based on the energy profile calculated in *Supplementary I*, the average velocity can be calculated with different width of contact. Ignoring the spacing between contacts and the skyrmion size change during the shift, the average velocity is calculated with varying the width of contact with fixed skyrmion radius 31 nm ($K = 310 \text{ kJ/m}^3$, $D = 1.2 \text{ mJ/m}^2$). The numerical result reveals that the maximum of the average velocity is obtained when the contact width $l$ is 58 nm, which is about twice of the skyrmion radius $R_{sk}$. This conclusion is coherent with the result in *Supplementary II*, as the energy platform will decrease the motion velocity.

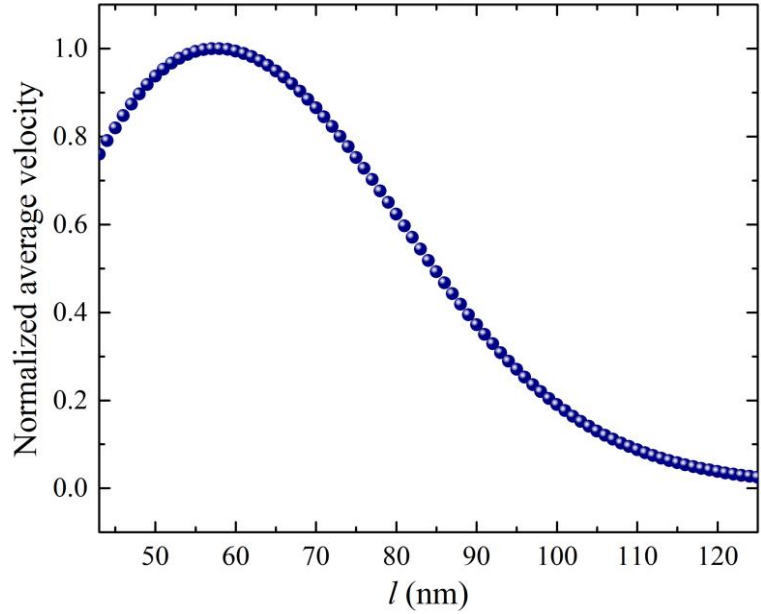

**Figure S4.** Velocity change (normalized) with different contact width $l$.

**Supplementary IV: Radius and velocity change with different magnetic parameters**

In this part, we change magnetic parameters $D$ and $K$ to study the impact on skyrmion radius and shift velocity $v_0$, while the contact width $l$ is fixed at 70 nm. Figure S4 shows the change of skyrmion radius $R$ and average velocity $v_0$ with different $D$ and $K$. $R$ increases when $D$ increases or $K$ decreases, which consists with equation (4). $v_0$ increases when $R$ is less than 35 nm and decreases when $R$ is larger than 35 nm, which is also coherent with the conclusion in *Supplementary III.*

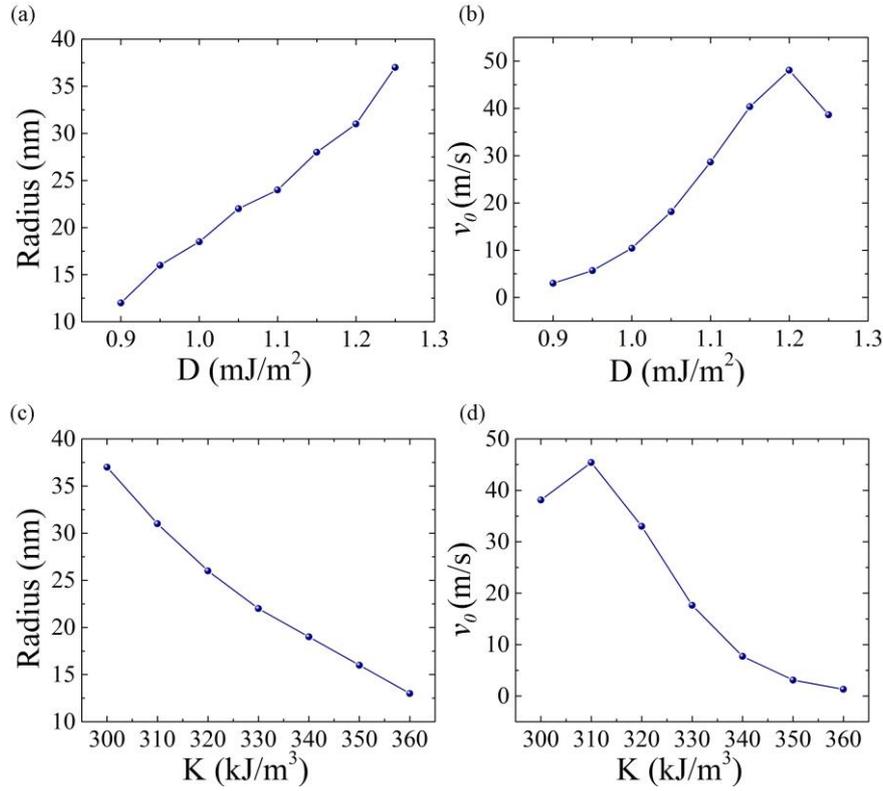

**Figure S6.** Radius and velocity change with different $D$ and $K$. (a) Radius increases with larger $D$. (b) Velocity increases when $D$ is less than $1.2\ \text{mJ}/m^2$ (radius about 35 nm) and decreases when $D$ is larger than $1.2\ \text{mJ}/m^2$. (c) Radius decreases with larger $K$. (d) Velocity increases when $K$ is less than $310\ \text{kJ}/m^2$ (radius about 35 nm) and decreases when $K$ is larger than $310\ \text{kJ}/m^2$. For (a) and (b), $K$ is fixed at $310\ \text{kJ}/m^2$. For (c) and (d), $D$ is fixed at $1.2\ \text{mJ}/m^2$.

## Supplementary V: Energy consumption

In our system, the energy dissipation is mainly from the charge of capacitive dielectric layer, namely the MgO layer. To achieve the skyrmion velocity around 100 m/s, 0.15 V voltage is applied on the contacts. Taking 1 nm MgO as the gate oxide with the relative permittivity of 7 and an area of 60 nm×100 nm, the energy required is 8.4 aJ.

Compared with the reported results from experiment for the same velocity, the required current density is around $5 \times 10^{11}\ \text{A} \cdot \text{m}^{-2}$.[3,4] According to the experimental work, the film structure is a [Pt(4.5 nm)/CoFeB(0.7 nm)/MgO(1.4 nm)]$_{15}$ multilayer stack. The resistivity of metals is taken as $1 \times 10^{-7}$ Ω·m. In this case, to drive one skyrmion shift of 70 nm, the Joule heating is 8.2 fJ. Thus, the energy dissipation in our voltage-driven method is three orders lower than that in current-driven skyrmion motion.

**Supplementary VI: Video of the motion**

Two voltage-driven motion videos are attached. Both of them contains three shifts. The first one is under 0.3 V voltage and the average velocity is about 100 m/s. The second one is under 0.45 V voltage and the average velocity is about 249 m/s.

**REFERENCES**

1. Romming, N.; Kubetzka, A.; Hanneken, C.; von Bergmann, K.; Wiesendanger, R. Field-Dependent Size and Shape of Single Magnetic Skyrmions. *Phys. Rev. Lett.* **2015**, *114*, 177203.

2. Wang, X. S.; Yuan, H. Y.; Wang, X. R. A Theory on Skyrmion Size and Profile. *ArXiv Preprint* **2018**, arXiv:1801.01745v1.

3. Fert, A.; Reyren, N.; Cros, V. Magnetic Skyrmions: Advances in Physics and Potential Applications. *Nat. Rev. Mater.* **2017**, *2*, 17031.

4. Woo, S.; Litzius, K.; Krüger, B.; Im, M.-Y.; Caretta, L.; Richter, K.; Mann, M.; Krone, A.; Reeve, R. M.; Weigand, M.; *et al.* Observation of Room-Temperature Magnetic Skyrmions and Their Current-Driven Dynamics in Ultrathin Metallic Ferromagnets. *Nat. Mater.* **2016**, *15*, 501–506.